\documentclass[12pt]{iopart}
\usepackage{mathrsfs}

\expandafter\let\csname equation*\endcsname\relax
\expandafter\let\csname endequation*\endcsname\relax
\usepackage{amsmath, amssymb}
\usepackage{graphicx,color}
\usepackage{lscape}
\usepackage{color}

\def\puni{\varphi^{\rm uni}}
\def\pssd{\varphi^{\rm ssd}}
\def\Puni{\Phi^{\rm uni}}
\def\Pssd{\Phi^{\rm ssd}}

\begin{document}

\title[]{
Sine-square deformation and supersymmetric quantum mechanics
}

\author{Kouichi Okunishi}
\address{Department of Physics, Faculty of Science, Niigata University, Japan}
\author{Hosho Katsura}
\address{Department of Physics, Graduate School of Science, University of Tokyo, Japan }

\date{\today}

\begin{abstract}
We investigate the sine-square deformation (SSD) of free fermions in one-dimensional continuous space.
On the basis of supersymmetric quantum mechanics, we prove the correspondence between the many-body ground state of the system with SSD and that of the uniform system with periodic boundary conditions. 
We also discuss the connection between the SSD in the continuous space and its lattice version, where the geometric correction due to the real-space deformation plays an important role in relating the eigenstates of the lattice SSD with those of the continuous SSD. 
\end{abstract}


\submitto{\JPA}

\section{Introduction}

Recently, spatial deformations of interaction couplings in low-dimensional quantum many-body systems have attracted much attention \cite{GendiarKN2009, HikiharaN2011, GendiarDLN2011, Katsura2011, Katsura2011B,ShibataH2011, Hotta2, Hotta, Hikihara_Suzuki, Tada:2014kza, Ishibashi, MaruyamaKH2011,exponential, rainbow, cosh}.
A smooth cutoff of the coupling encoded in the real space is generally expected to suppress the scattering at the boundaries~\cite{VekicW1993}. 
Among the various schemes, the sine-square deformation (SSD) in one-dimensional (1D) quantum critical systems is of particular interest, because the many-body ground state of a system with SSD almost coincides with that of the uniform system with periodic boundary conditions (PBC).  
This is due to the nearly exact cancellation of the boundary scattering in the system with SSD~\cite{GendiarKN2009, HikiharaN2011}. 
In a class of 1D critical systems which are reducible to free fermions, moreover, the ground-state correspondence can be made exact. 
In fact, it was proved that the ground state of the spin-$1/2$ XY chain with SSD is identical to that of the uniform chain with PBC~\cite{Katsura2011}. 
The same holds true for the tight-binding chains~\cite{MaruyamaKH2011} and transverse field Ising chains at criticality~\cite{Katsura2011B}. 
More generally, the ground-state correspondence between the SSD and uniform systems can be explained in terms of conformal field theory (CFT), where the Virasoro generator $L_{\pm 1} $ identified with the chiral part of the Hamiltonian of the SSD system~\cite{chiral} annihilates the CFT vacuum~\cite{Katsura2011B}.
Another interesting aspect of the SSD is that it has a variety of applications;
For instance, the nature of the quasi-localized edge excitations originating from the SSD is essential in a `grand-canonical' approach for finite-size systems, where smooth magnetization curves of quantum spin systems can be obtained with very high accuracy~\cite{Hotta2, Hotta}.
Besides, very recently, a further relation between the SSD and CFT, and its relevance to string theory is also pointed out in Refs.~\cite{Tada:2014kza, Ishibashi}

A key to understanding the ground-state correspondence between the SSD and uniform systems is that each chiral part of the SSD Hamiltonian annihilates the many-body ground state of the uniform system with PBC~\cite{MaruyamaKH2011}. 
Here, we remark that the plane-wave representation of the single-particle basis, which is a suitable description for the uniform system with PBC,  plays a very essential role in verifying the above statement.
However, as shown numerically in Ref.~\cite{ShibataH2011,Hotta2,Hotta}, low-lying single-particle states in the SSD system seem to be localized 
around the center or edges of the chain,  which would be rather puzzling. 
A natural question is: ``What is the unified view of the physics behind the plane-wave description and the single-particle eigenstates of the SSD?"
In this paper, we explore this question by examining the SSD of the free-fermion system defined in 1D continuous space. 
To this end, we formulate the SSD problem as the inverse problem of the quantum mechanical problem of a particle in the inverse sine-square ($1/\sin^2$) potential. 
An elegant way to solve the direct problem is to use supersymmetric (SUSY) quantum mechanics with the shape invariance~\cite{Witten,SUSYQM}. 
With this, we show that a set of the single-particle eigenstates of the SSD problem can be obtained in closed form for specific values of the parameter $\mu$, which will be introduced in the next section.
Then we show that the Slater determinant obtained by filling these single-particle states up to the zero energy is identical to the Slater determinant corresponding to the ground state of the uniform system with PBC. 
Moreover, we discuss the relation of the SSD in a lattice with that in the continuous space, and clarify the role of the geometric correction involved in the single-particle wavefunctions of the SSD. 
Then, we demonstrate that a number of properties of the lattice SSD found in the previous study~\cite{Hotta} can be well explained on the basis of the SSD wavefunctions obtained with the use of SUSY quantum mechanics.

The rest of the paper is organized as follows.
In section 2, we define the SSD problem of free fermions in 1D continuous space.
In section 3, we formulate the single-particle problem with SSD as the inverse problem of the $1/\sin^2$ potential problem. 
Then we obtain the single-particle eigenstates for specific values of the chemical potentials. 
In section 4, we prove the ground-state correspondence between the uniform and SSD systems by showing that the single-particle eigenstates for the SSD can be expressed as linear combinations of plane waves. 
In section 5, we briefly comment on the single-particle states with positive energy in the SSD system. 
In section 6, we compare the SSD in the continuous space with that in a lattice, and discuss the role of the geometric correction. 
In the last section, we summarize the results obtained and discuss the prospects for future researches.

\section{Definition of the SSD problem in a one-dimensional ring}
\label{sec:problem_setup}


We start with the definition of the SSD problem in 1D continuous space. 
Consider $N$ fermions on a ring of length $L$. 
We denote by ${\hat c}^\dagger (x)$ and ${\hat c}(x)$ the fermionic field operators at position $x$ ($0 \le x \le L$). 
They satisfy the canonical anticommutation relations: 
$\{ {\hat c}(x), {\hat c}^\dagger (y) \} = \delta (x-y)$, 
$\{ {\hat c}(x), {\hat c} (y) \} =\{ {\hat c}^\dagger (x), {\hat c}^\dagger (y) \}=0$. 
In analogy with lattice systems, the Hamiltonian for the system with a generic deformation is defined as
\begin{equation}
{\hat {\cal H}}_{\rm cont} = \int^L_0 \, dx\, {\hat c}^\dagger (x)\,
				f(x) \left( -\frac{d^2}{dx^2} - \mu \right) {\hat c}(x) 
\label{Hamiltonian}
\end{equation}
where $\mu$ is the ``chemical potential" introduced for later convenience, and $f(x)$ encodes the spatially varying couplings. 
Here we set $\hbar=2m=1$.  
If $f(x)=1$, the systems is nothing but the uniform ring with non-relativistic free fermions.
For the case of SSD, $f(x)$ is defined as 
\begin{equation}
f(x) := \sin^2 \left(\frac{\pi }{ L} x\right).
\end{equation}
Thus, the energy scale of the system smoothly approaches zero as $x \to 0$ and $x \to L$, implying that the ring is effectively disconnected at $x=0$ (mod $L$).
In this sense, the boundary conditions of the SSD problem can be interpreted as open ones.

A general $N$-particle eigenstate of ${\hat {\cal H}}_{\rm cont}$ is of the form:
\begin{equation}
|\Phi \rangle = \int^L_0 dx_1 \cdots \int^L_0 dx_N\, 
{\hat c}^\dagger (x_1) \cdots {\hat c}^\dagger (x_N) |0\rangle 
\Phi (x_1, ..., x_N)\, ,
\end{equation}
where $|0\rangle$ is the state with no fermions. 
Since the Hamiltonian is quadratic in ${\hat c}$ and ${\hat c}^\dagger$, the $N$-body wavefunction in the coordinate representation $\Phi (x_1, ..., x_N)$ can be constructed as a determinant of single-particle wavefunctions $\varphi (x)$ which are determined by the Schr\"odinger-like equation:
\begin{equation}
-f(x) \left( \frac{d^2}{dx^2} +\mu \right)\varphi(x) = \xi \varphi(x),
\label{ssdh}
\end{equation}
where $\xi$ denotes the eigenvalue. 
Note that the Hamiltonian here is not self-adjoint in the standard sense and the appropriate inner product for the SSD problem will be introduced in section 5. 
For later convenience, we assume either periodic or antiperiodic boundary condition for $\varphi (x)$, depending on the total number of fermions: 
$\varphi (x+L) = (-1)^N \varphi (x)$. 
We denote by $\{\puni_n(x)\}$ and $\{\pssd_n(x)\}$ the sets of single-particle eigenfunctions of the uniform ($f(x)=1$) and SSD ($f(x)=\sin^2 \frac{\pi}{L}x$) systems, respectively. 
For the uniform system, $\{\puni_n(x)\}$ is of course described by the plane waves and is easy to obtain. 
For periodic boundary conditions, one finds that the linear space of $\{ \puni_n (x) \}$ is spanned by
\begin{equation}
\exp \left( \pm i \frac{2 n \pi}{L}x \right), 
\quad (n=0,1,2, ...),
\end{equation}
and for antiperiodic boundary conditions,
\begin{equation}
\exp \left( \pm i \frac{(2 n+1) \pi}{L}x \right), 
\quad (n=0,1,2, ...).
\end{equation}
In contrast, to obtain $\{ \pssd_n (x) \}$ for the SSD problem is an intriguing problem, because the translational symmetry is explicitly broken by the position-dependent factor $f(x)$. 
However, for specific values of the parameter $\mu$, one can obtain them in closed form by exploiting the hidden SUSY in the problem.

\section{Single-particle eigenstates of the SSD problem}
\label{sec:susy_solution}

In this section, we present a systematic procedure to construct single-particle eigenstates of the SSD problem. 
To this end, we introduce the following parametrization of $\xi$:
\begin{equation}
\xi = - \left(\frac{\pi}{L}\right)^2\beta (\beta-1),
\label{SSDspectrum}
\end{equation}
and rewrite Eq. (\ref{ssdh}) with $f(x)=\sin^2 \frac{\pi}{L}x$ as
\begin{equation}
H_\beta\, \varphi (x) = \mu \varphi (x),
\label{SSDpotential}
\end{equation}
where the ``Hamiltonian" $H_\beta$ is defined as
\begin{equation}
H_\beta = - \frac{d^2}{dx^2} + \left(\frac{\pi}{L}\right)^2 
\frac{\beta (\beta-1)}{\sin^2 \frac{\pi}{L} x }. 
\end{equation}
In the standard setup, ``energy eigenvalue" $\mu$ is determined for a given parameter $\xi$ (or, equivalently, $\beta$), where the corresponding eigenfunction has a wavepacket-like shape. 
On the other hand, however, what we need here is to determine the energy eigenvalue $\xi$ for a given chemical potential $\mu$. In this sense, the SSD problem can be formulated as the inverse problem of the $1/\sin^2$ potential problem.

It is known that the $1/\sin^2$ potential problem can be solved algebraically since the system has SUSY and shape invariance~\cite{SUSYQM}. 
This property is also deeply related to the integrable many-body system with an inverse sine-square interaction solved by Sutherland~\cite{Sutherland, Kuramoto}.
Here, we briefly summarize the SUSY quantum mechanics for the $1/\sin^2 $ potential problem. 
The eigenvalue problem of Eq. (\ref{SSDpotential}) can be solved algebraically for non-negative integers $\beta=0,1,2,3 \cdots$.
Let us start with defining intertwiners
\begin{equation}
A_\beta = \frac{d}{dx} + W'_\beta(x),  
\quad  A^\dagger_\beta = -\frac{d}{dx} +W'_\beta(x)
\end{equation}
with the superpotential $W_\beta(x) := - \beta \ln \left( \sin \frac{\pi}{L}x \right)$. 
Using these operators, 
the Hamiltonian $H_\beta$ in Eq. (\ref{SSDpotential}) can be written as 
\begin{eqnarray}
H_\beta = A_\beta^\dagger A_\beta + \left(\frac{\pi}{L}\right)^2\beta^2.  
\label{eq:quad}
\end{eqnarray}
According to SUSY quantum mechanics, it is well-known that  $A^{}_\beta A^\dagger_\beta$ and $A^\dagger_\beta A^{}_\beta$ form the SUSY-partner Hamiltonians, and the shape invariance ensures the recursive relation  
\begin{equation}
A^{}_\beta A_\beta^\dagger = A^\dagger_{\beta+1} A_{\beta+1}^{} + \left(\frac{\pi}{L}\right)^2 [(\beta+1)^2- \beta^2]
=  H_{\beta+1} - \left(\frac{\pi}{L}\right)^2\beta^2.
\label{eq:recursion}
\end{equation}
In addition, it is easily confirmed that $A_\beta \psi^0_\beta(x)=0$ with $\psi^0_\beta(x) := (\sin \frac{\pi}{L} x )^\beta$, which corresponds to the eigenfunction of $H_\beta$ with the lowest eigenvalue. 
Note that Eq. (\ref{eq:quad}) implies that the eigenvalue of $H_\beta$ is bounded from below by $(\frac{\pi}{L})^2 \beta^2$. 
It follows from Eq. (\ref{eq:recursion}) that, for a given non-negative integer $\beta$, the $\ell$th eigenstate of $H_\beta$ is constructed as
\begin{equation}
\psi_\beta^\ell(x) = A^\dagger_\beta \cdots A^\dagger_{\beta+\ell-2}A^\dagger_{\beta+\ell-1}\psi^0_{\beta+\ell}(x).
\label{A_state}
\end{equation}
The explicit form of the eigenfunction can be written as 
\begin{equation}
\psi_\beta^\ell(x) ={\cal N}^\ell_\beta\, C_\ell^\beta\left(\cos \frac{\pi}{L}x \right)\psi_\beta^0(x)
\label{gegenbauer}
\end{equation}
with the coefficient
\begin{equation}
{\cal N}^\ell_\beta = \left( -\frac{\pi}{L} \right)^\ell 
\frac{\ell!\, (2\beta+2\ell-1)!\, (\beta-1)! }
{2^\ell\, (2\beta+\ell-1)!\, (\beta+\ell-1)!},
\end{equation}
where $C_\ell^\beta (z)$ denotes the Gegenbauer polynomial of the $\ell$th order~\cite{tableofintegral,Kuramoto}. 
The coefficient ${\cal N}^\ell_\beta$ can be obtained by demanding that $A_\beta \psi^\ell_\beta (x) = \left( \frac{\pi}{L} \right)^2 \ell (\ell+2\beta) \psi^{\ell-1}_{\beta+1} (x)$ and the following recursion relation for the Gegenbauer polynomials are consistent:
\begin{equation}
\frac{d}{dz} C^\beta_\ell (z) = 2\beta\, C^{\beta+1}_{\ell-1} (z).
\end{equation}
Then, the corresponding eigenvalue of $H_\beta$ is obtained as 
\begin{equation}
\mu= \left( \frac{\pi}{L} \right)^2(\beta+\ell)^2,
\label{chem_p}
\end{equation}
for $\ell=0,1,2, \cdots$.
Thus, if the sum of $\beta$ and $\ell$ is constant, say $\beta+\ell=q$,  then $\mu$ takes the same value among $\psi^0_q (x), \psi^1_{q-1}(x), \cdots, \psi^{q-\beta}_{\beta}(x), \cdots, \psi^q_0 (x)$. 

Here, we list in Table I the first few eigenfunctions of $H_\beta$ with $\beta=0, 1, ..., 5$, from which the relation between the $1/\sin^2$-potential problem and the SSD problem can be seen clearly. 
On the one hand, one finds the eigenfunctions of $H_\beta$ in the column labeled by $\beta$. 
On the other hand, the functions in the row labeled by $q$ give the solutions of the inverse problem (up to unimportant overall factors). 
For a particular chemical potential parameterized by $\mu= (\frac{\pi}{L})^2 q^2$, we obtain them as $\psi^{q-\beta}_{\beta} (x)$ ($\beta = 0, 1, \dots, q$) with the corresponding energy eigenvalues $\xi=-\left(\frac{\pi}{L}\right)^2\beta(\beta-1)$.
Note that $\psi^0_q (x)$ can be thought of as the single-particle ground state of the SSD problem, while $\psi^{q}_{0}(x)$ and $\psi^{q-1}_{1} (x)$ are the plane-wave solutions with zero energy. 

\begin{landscape}
\begin{table*}
\begin{tabular}{c|c|cc|lllll} 
 B.C. & $q\backslash\beta$ & 0&1 &2 &3&4&5& \\ \hline
 AP &\ 5 &$\cos\frac{5 \pi x}{L}$ & $\sin\frac{5 \pi x}{L}$ & $\sin^2\frac{\pi x}{L}\cos\frac{\pi x}{L}(12\cos^2\frac{\pi x}{L}-5)$&$\sin^3\frac{\pi x}{L}(8\cos^2\frac{\pi x}{L}-1) $&$\sin^4\frac{\pi x}{L}\cos\frac{\pi x}{L} $& $\sin^5\frac{\pi x}{L}$& \\
 P & 4 &$\cos\frac{4 \pi x}{L}$ & $\sin\frac{4 \pi x}{L}$ & $\sin^2\frac{\pi x}{L}(6\cos^2\frac{\pi x}{L}-1) $&$\sin^3\frac{\pi x}{L}\cos\frac{\pi x}{L} $&$\sin^4\frac{\pi x}{L} $&& \\
 AP & 3 &$\cos\frac{3 \pi x}{L}$ & $\sin\frac{3 \pi x}{L}$ & $\sin^2\frac{\pi x}{L}\cos\frac{\pi x}{L} $&$\sin^3\frac{\pi x}{L} $&&& \\
 P & 2 &$\cos\frac{2 \pi x}{L}$ & $\sin\frac{2 \pi x}{L}$ & $\sin^2\frac{\pi x}{L} $&&&& \\
 AP & 1 & $\cos\frac{\pi x}{L}$ & $\sin\frac{\pi x}{L}$ & &&&& \\
 P & 0 &const. &  & &&&& \\ \hline
\end{tabular}
\caption{List of eigenfunctions of $H_\beta$ up to $\beta=5$. 
P and AP in the B.C. column, respectively, indicate the periodic and antiperiodic boundary condition imposed on the eigenfunctions. 
The columns for $\beta=0$ and $1$ correspond to the plane-wave solutions of the uniform system. 
The functions in each row give the eigenfunctions of the SSD problem. 
For fixed $q$, they are identical to 
$\psi^0_q (x), \psi^1_{q-1}(x), \cdots, \psi^{q-\beta}_{\beta} (x), \cdots, \psi^q_0 (x)$
up to unimportant overall factors. 
}
\end{table*}
\end{landscape}

In order to precisely see the correspondence between the solutions of the $1/\sin^2$-potential and the SSD problems, we further comment on the boundary conditions. 
In the literature, it is often assumed that $\beta \ge 1$ in Eq. (\ref{SSDpotential}). 
Here, we have added the SUSY partners for $\beta=0$ and $1$. 
For both $\beta=0$ and 1, there is no potential term in $H_\beta$ and thus the system is uniform in $0 \le x \le L$. 
However, an important difference between them is that the boundary conditions for $\beta=0$ are Neumann, i.e., $\varphi'(0)=\varphi'(L)=0$, in contrast to the Dirichret ones for $\beta=1$.
Thus, the eigenfunctions for $\beta=0$ and $1$ are described by cosine and sine functions, respectively.
In terms of SUSY, the lowest eigenstate of $H_0$ is a singlet with zero energy, because it is annihilated by both $A_0$ and $A^\dagger_0$. 
Any other eigenstate of $H_0$ has the corresponding eigenstate of $H_1$, which give rise to a doublet of eigenstates with the same eigenvalue. 
Thus, we can express the plane-wave solutions for a ring of length $L$ by superposing the eigenfunctions of $H_0$ and $H_1$. 
We recall here that we have imposed periodic (antiperiodic) boundary conditions on $\varphi (x)$ depending on the total number of fermions $N$. 
Therefore, in Table I we should take the eigenfunctions in the row labeled by even (odd) $q$ when $N$ is even (odd). 
We also note that the boundary conditions for the eigenfunctions are indicated by P (periodic) and AP (antiperiodic) in Table I. 
In contrast to the cases of $\beta=0$ and $1$, the eigenfunctions (\ref{gegenbauer}) with $\beta \ge 2$ enjoy $\varphi(0)=\varphi(L)=\varphi'(0)=\varphi'(L)=0$, which reflects the fact that they have wavepacket-like shapes and are disconnected at $x=0$ (mod $L$).

\section{Equivalence of two Slater determinants}
\label{sec: equivalence}

In this section, we give a proof that the Slater determinant of the plane waves, corresponding to the ground state of the uniform system with PBC, is identical to the Slater determinant of $\psi^{q-\beta}_\beta (x)$, which are the eigenfunctions of the SSD problem. 
This can be thought of as a continuous version of the ground-state correspondence proved in a class of 1D lattice models reducible to free fermions~\cite{Katsura2011, Katsura2011B, MaruyamaKH2011}. 
In the following, we focus on the case where the total fermion number $N$ is even. It is, however, very straightforward to extend the proof to the case of odd $N$. 

Let us first consider the many-body ground state of $N$ (even) fermions on the uniform periodic ring with $\mu = (\frac{\pi}{L})^2 N^2$, which is the Slater determinant obtained by filling the plane-wave states up to $\xi=0$. 
It is easy to see from Table I that the plane-wave states with even $q$ can be obtained as linear combinations of eigenfunctions in the columns labeled by $\beta=0$ and $1$, which satisfy the PBC. 
Since $\mu = (\frac{\pi}{L})^2 N^2$, one can easily confirm that the states in the $q=N$ row with $\beta=0$ and $1$ are just located at $\xi=0$.  
Thus, the many-body ground state is two-fold degenerate, because either one of the two $\xi=0$ states is occupied by a fermion. 
In the following, we assume that $\beta=1$ state is occupied. 
Then, the single-particle states involved in the Slater determinant is written as
\begin{eqnarray}
\puni_n (x) = \left\{
\begin{array}{cc}
\cos \frac{2n \pi}{L}x \quad & (n=0, 1, ..., \frac{N}{2}-1), \\
 & \\
\sin \frac{(2n-N+2)\pi}{L}x \quad & (n=\frac{N}{2}, \frac{N}{2}+1, ..., N-1),
\end{array}
\right.
\end{eqnarray}
where $\cos \frac{N\pi}{L}x$ does not appear because we have assumed that $\beta=0$ state is empty. 
Note that, if we consider the case of odd $N$, the boundary conditions are antiperiodic, where we should take the wavefunctions only from $q=$odd sectors.
In terms of $\puni_n (x)$, the $N$-fermion ground-state wavefunction is written as 
\begin{equation}
\Puni (x_1, ... , x_N) = \det_{1 \le i, j \le N} \, \left[ \puni_{i-1} (x_j) \right].
\label{SD1}
\end{equation}

We next consider a many-body ground state of the SSD problem with the same setup, i.e., $N$ is even and $\mu= (\frac{\pi}{L})^2N^2$. 
As shown in the previous section, the single-particle eigenfunctions of the SSD problem is given by
\begin{equation}
\pssd_{\beta} (x) = \psi^{N-\beta}_{\beta} (x) \quad (\beta=0,1,2, ..., N),
\end{equation}
where $\psi^{N-n}_{n} (x)$ is defined in Eq. (\ref{A_state}). 
The many-body ground state is then described by the Slater determinant of $\psi_\beta^{N-\beta} (x)$ ($1\le \beta \le N$), which span the subspace of $\{ \pssd_n (x) \}^N_{n=0}$. 
Here, we again assume that not $\beta=0$, but $\beta=1$ state with $\xi=0$ is occupied by a fermion. 
Then, the $N$-fermion ground-state wavefunction is given by
\begin{equation}
\Pssd (x_1, ..., x_N) = \det_{1 \le i,j \le N}\, \left[ 
\pssd_{i} (x_j) \right].
\label{SD2}
\end{equation}

At first sight, one might think that the Slater determinants Eqs. (\ref{SD1}) and (\ref{SD2}) are not related to each other because $\{ \puni_n (x) \}$ and $\{ \pssd_n (x) \}$ look completely different. 
Nevertheless, we will show that the following relation
\begin{equation}
\Pssd (x_1, ..., x_N) = C_N\, \Puni (x_1, ..., x_N)
\label{eq:theorem}
\end{equation}
holds for any $N$ (even), where $C_N$ is a constant independent of $x_1, ..., x_N$. This means that the many-body ground states of the uniform and SSD systems are identical to each other. 
The above relation can be proved by noting that the two Slater determinants become equivalent if there exists a linear transformation between $\{ \puni_n (x) \}^{N-1}_{n=0}$ and $\{ \pssd_n (x) \}^{N}_{n=1}$ that is non-singular, i.e., the determinant of its transformation matrix is nonzero. 
For this purpose, however, the eigenfunctions in terms of the Gegenbauer polynomials (\ref{gegenbauer}) are not so useful.
In the following, we will consider a more direct treatment of $\psi_\beta^{N-\beta}(x)$. 
Let us rewrite the intertwiner $A^\dagger_\beta$ as 
\begin{equation}
A^\dagger_\beta  = -\sin^{-\beta} (\tilde{x}) \frac{d}{dx} \sin^\beta (\tilde{x}),
\end{equation}
where we have introduced the variable $\tilde{x} := \frac{\pi}{L} x $ for simplicity.
This leads us to the Rodrigues formula for the eigenfunctions Eq. (\ref{A_state}),
\begin{equation}
\psi_\beta^{N-\beta}(x)= \sin^{-\beta +1 } \tilde{x}   \left( \frac{-1}{\sin \tilde{x}} \frac{d}{dx} \right)^{N-\beta} \sin^{2N-1} \tilde{x}. 
\label{eq:Rod}
\end{equation}
This implies that the single-particle states can be described in terms of only trigonometric functions of ${\tilde x}$. 
Note that the explicit form depends on whether $N$ and $\beta$ are even or odd. Here, recall that $N$ should be even for the PBC. 
We can thus write $N= 2\theta $ with a positive integer $\theta$. 
For $\beta=2\nu$ ($\nu=0, 1, ..., \theta$), we get from Eq. (\ref{eq:Rod}) the following expansion: 
\begin{equation}
u_\beta(x) :=(-1)^{\theta-\nu}\left(\frac{\pi}{L}\right)^{-2(\theta-\nu)} \frac{(2\theta+2\nu-1)!}{(4\theta-1)!}\psi^{2(\theta-\nu)}_{2 \nu}(x) 
= \sum_{n=\nu}^\theta  a_{2\nu,2n} \sin^{2n} (\tilde{x}),
\label{eq:expansion1}
\end{equation}
where
\begin{equation}
a_{2\nu,2n} = \prod_{k=n}^{\theta-1}\frac{(2k+1)(k+1)-\nu(2\nu-1)}{2(k - \theta)(k+\theta)}
\end{equation}
with $a_{2\nu,2\theta} := 1$.
For $\beta=2\nu-1$ ($\nu=1,2, ..., \theta$), we also obtain another linearly-independent solution of odd parity as
\begin{eqnarray}
v_\beta(x) :=&& (-1)^{\theta-\nu+1}\left(\frac{\pi}{L}\right)^{-2(\theta-\nu)-1} \frac{(2\theta+2\nu-2)!}{(4\theta-1)!}\psi^{2(\theta-\nu)+1}_{2\nu-1}(x) \nonumber \\
&&= \cos (\tilde{x}) \sum_{n=\nu}^{\theta}  a_{2\nu-1,2n-1} \sin^{2n-1} (\tilde{x}),
\label{eq:expansion2}
\end{eqnarray}
where
\begin{equation}
a_{2\nu-1,2n-1} = \prod_{k=n}^{\theta-1}\frac{k(2k+1)-(2\nu-1)(\nu-1)}{2(k - \theta)(k+\theta)},
\end{equation}
with $a_{2\nu-1,2\theta-1}:= 1$.

Since $\psi^{N-2\nu}_{2\nu}(x)$ ($\psi^{N-2\nu+1}_{2\nu-1}(x)$) and $u_{2\nu}(x)$ ($v_{2\nu-1}(x)$) are the same up to an overall constant, the only thing remaining to prove is to show that there is a linear transformation between the spaces of 
$\{ u_\beta(x), v_\beta(x)\}$ and $\{ \puni_n (x) \}^{N}_{n=0}$
such that it is invertible. From Eqs. (\ref{eq:expansion1}) and (\ref{eq:expansion2}), one finds the following relations:
\begin{eqnarray}
\begin{pmatrix}
u_0(x)                \\
u_2(x)                \\
 \vdots            \\
u_N(x)                
\end{pmatrix}
&=&
\begin{pmatrix}
a_{0, 0} &  a_{0,2}&  \cdots & a_{0,N}  \\
        &  a_{2,2}&   \cdots      & a_{2,N}  \\
        &        &   \ddots&   \vdots   \\
        &        &         & a_{N,N}  
\end{pmatrix}
\begin{pmatrix}
  1            \\
 \sin^2(\tilde{x})  \\
 \vdots            \\
 \sin^N(\tilde{x})  
\end{pmatrix},
\label{vectorspaceu}\\
\begin{pmatrix}
v_1(x)                \\
v_3(x)                \\
 \vdots            \\
v_{N-1}(x)                
\end{pmatrix}
&=&
\begin{pmatrix}
a_{1, 1} &  a_{1,3}&  \cdots & a_{1,N-1}  \\
        &  a_{3,3}&    \cdots     & a_{3,N-1}  \\
        &        &   \ddots&   \vdots   \\
        &        &         & a_{N-1,N-1}  
\end{pmatrix}
\begin{pmatrix}
 \cos(\tilde{x}) \sin(\tilde{x})                \\
 \cos(\tilde{x}) \sin^3(\tilde{x})  \\
 \vdots            \\
 \cos(\tilde{x}) \sin^{N-1}(\tilde{x})  
\end{pmatrix},
\end{eqnarray}
where matrix elements which are zero are left empty. 
These upper triangular matrices are non-singular and invertible since the diagonal element $a_{\beta, \beta}$ is nonzero for all $\beta$. 
Thus, the linear space of the SSD eigenfunctions is equivalent to that of $\{\sin^{2n}(\tilde{x}) \}$ and $\{\cos(\tilde{x})\sin^{2n-1}(\tilde{x}) \}$.
According to the standard formula of the trigonometric functions~\cite{sinexpansion}, moreover, the linear spaces spanned by $\{\sin^{2n}(\tilde{x}) \}$ and $\{\cos(\tilde{x})\sin^{2n-1}(\tilde{x}) \}$ are respectively equivalent to those by $\{\cos(2n\tilde{x})\}$ 
and $\{\sin(2n\tilde{x})\}$.
Hence, the linear space of $\{ \pssd_n (x) \}^{N}_{n=0}$ can be spanned by the basis elements of $\{\cos(2n\tilde{x})\}^{N/2}_{n=0}$ and those of $\{\sin(2n\tilde{x})\}^{N/2}_{n=1}$, indicating that the single-particle basis of the SSD problem is identical to that of the plane waves. 

Here, we should recall that the $\xi=0$ state in the $\beta=0$ sector is assumed to be empty of fermion, which implies that $u_0(x)(\propto \psi_0^{N}(x))$ is not included in $\{ \pssd_n (x) \}^{N}_{n=1}$.
However, we can easily check  $u_0(x)=\sum_{n=0}^{N/2} a_{0,2n} \sin^n(\tilde{x})\propto \cos(N\tilde{x})$~\cite{redundant}.
Thus, the removal of $u_0(x)$ from the SSD basis results in the linear space $\{ \puni_n (x) \}^{N-1}_{n=0}$, which is identical to that of the plane waves below the Fermi level in the uniform system. 
Thus, we have obtained the desired relation Eq. (\ref{eq:theorem})

If the total fermion number $N$ is odd,  we should take eigenfunctions in the row labeled by $q=N$, which are antiperiodic in $x$.  
The proof proceeds along the same lines as that of the $N=$even case, and will not be given here.

\section{Single-particle states with positive energy}

The equivalence between the uniform ring and the SSD problem is established only for the ground states of many fermions. 
Here, we comment on single-particle excited states above the Fermi level, i.e., the positive energy ($\xi >0$) solutions of the SSD problem. 
From the correspondence between the SSD and inverse sine-square problems, one can see that they translate into the eigenfunctions of the attractive $1/\sin^2$ potential problem. 
As noted in Appendix A, the solution of the $1/\sin^2$  potential problem with the attractive coupling is classified into two cases: (i) $0<\xi <\frac{1}{4} (\frac{\pi}{L})^2 $ and (ii) $\frac{1}{4} (\frac{\pi}{L})^2 <\xi$. 
The case (i) corresponds to Eq. (\ref{SSDpotential}) with $0<\beta<1$. 
For a non-integer value of $\beta$, however,  the solution of Eq. (\ref{SSDpotential}) is given by an analytic continuation of the Gegenbauer polynomial, which exhibits a singular behavior at the boundaries.
Thus, the physical solution is permitted only in the limit of $\beta\to 0$ or $1$, which is reduced to the plane-wave solution of $\beta=0$ or 1. Recalling that the boundary condition for $\beta=0$ is Neumann and that for $\beta=1$ is the Dirichlet, one may think of the solution for $\beta=0$ as the zero-energy limit of a scattering state, while that for $\beta=1$ as the zero-energy limit of a bound state. 
For the case (ii), the attractive potential problem is ill-defined, where the energy of the single-particle ground state (in the sense of the potential problem) collapses to $-\infty$~\cite{Landau,Sutherland}.
Thus, the SSD problem does not have a proper single-particle excited state with positive energy, unlike the SSD for the lattice systems.
We conclude that the SSD problem in continuous space is well defined only below the Fermi level.

\section{Geometric factor and lattice SSD}

In this section, we discuss the relation between the above continuous-space SSD and its lattice version, in the former of which a geometric factor associated with the inner product of the SSD eigenfunctions is of particular importance.
So far, we have not defined the inner product of the SSD eigenfunctions. 
This is because we should pay a particular attention to the geometric factor in the continuous SSD, unlike the SSD for lattice systems.
For instance, the standard definition of the inner product for the plane waves fails to satisfy the orthogonality relation among $\psi^{N-\beta}_\beta(x)$ ($\beta=1,2, ..., N$), though the linear space of them is identical to that of the plane waves.

For the continuous SSD, the self-adjoint form of the differential equation tells us that an appropriate inner product for $\psi^{N-\beta}_\beta(x)$ is
\begin{equation}
\langle\langle \psi^{N-\beta}_\beta,\, \psi^{N-\beta'}_{\beta'} \rangle\rangle := \frac{1}{L}  \int_0^L \frac{dx}{\sin^2\frac{\pi}{L}x}\,
\psi^{N-\beta}_\beta(x)\, \psi^{N-\beta'}_{\beta'}(x)  = I_{N,\beta}\, \delta_{\beta,\beta'},
\label{innerprod}
\end{equation}
where the coefficient $I_{N,\beta}$ is given by
\begin{equation}
I_{N,\beta} = 2^{2-2\beta}\, ({\cal N}^{N-\beta}_\beta)^2
\frac{(N+\beta-1)!}{(2\beta-1)\, (N-\beta)!\, [(\beta-1)!]^2},
\end{equation}
the derivation of which is given in \ref{app:coeffI}. 
The factor $\sin^2\frac{\pi}{L}x$ in the integrand in Eq. (\ref{innerprod}) can be regarded as a geometric factor attributed to the spatial deformation.
With this definition, the orthogonality of the single-particle wavefunctions for $\beta\ge 1$ is correctly recovered. 
Here, we should mention that  $\psi_0^N(x) \propto \cos(\frac{N\pi}{L}x )$ for $\beta=0$ is not normalizable with Eq. (\ref{innerprod}).
We will discuss this problem later. 

In order to see the role of the geometric factor, it is useful to consider the lattice version of the SSD problem, the Hamiltonian of which is defined as
\begin{eqnarray}
\!\!\!\!\!\!\!\!\!\!\!\!\!\!\!
{\hat {\cal H}}_{\rm lat}= -\sum_{n=1}^{L} \sin^2 \left(\frac{\pi}{L}n \right) ({\hat c}_n^\dagger {\hat c}_{n+1} + {\hat c}^\dagger_{n+1} {\hat c}_n ) 
-\mu\sum_{n=1}^{L} \sin^2 \left[ \frac{\pi}{L} \left( n-\frac{1}{2} \right) \right]{\hat c}^\dagger_n {\hat c}_n ,
\label{latticeSSD}
\end{eqnarray}
where ${\hat c}^\dagger_n$ and ${\hat c}_n$ are the creation and the annihilation operators, respectively, of a fermion at site $n$. 
The lattice spacing is set to unity so that the length of the chain $L$ is the same as the total number of sites.
We numerically solve the single-particle problem of ${\hat {\cal H}}_{\rm lat}$ and discuss its relation to the continuous SSD. 
Let $\chi_m (n)$ be the $m$th single-particle eigenfunction of ${\hat {\cal H}}_{\rm lat}$. 
The level index $m$ runs from $0$ to $L-1$.  
The inner product between $\chi_m (n)$ and $\chi_{m'}(n)$ is well-defined and the following orthogonality holds: $\langle \chi_m,\, \chi_{m'} \rangle := \sum^{L}_{n=1} \chi_{m}(n) \chi_{m'}(n) =\delta_{m,m'}$.  
This suggests that, unlike Eq. (\ref{innerprod}), the lattice SSD eigenfunctions do not involve any geometric factor. 
On the other hand, it is also expected that a naive continuum limit of the lattice SSD eigenfunction $\chi_m (n)$ may approach the eigenfunction of the continuous SSD problem $\psi_\beta^{N-\beta} (x)$. 
These seemingly contradictory facts can be reconciled by comparing the inner product for the continuous SSD with that for the lattice ones.  
This leads to the following correspondence:
\begin{equation}
\psi_\beta^{N-\beta}(x) \sim \sin \left( \frac{\pi}{L} n \right) \chi_m (n) ,
\label{CLcomparison}
\end{equation}
with $x\simeq n$, where $\sin ( \frac{\pi}{L} n)$ can be viewed as a geometric correction to $\chi_m$.   
For convenience, we fix the relation between $m$ and $\beta$ as $m = N-\beta$ so that the energy eigenvalues $\epsilon (m)$ of the lattice eigenfunctions are arranged in ascending order. 

\begin{figure}
\begin{center}
\includegraphics[width=7cm]{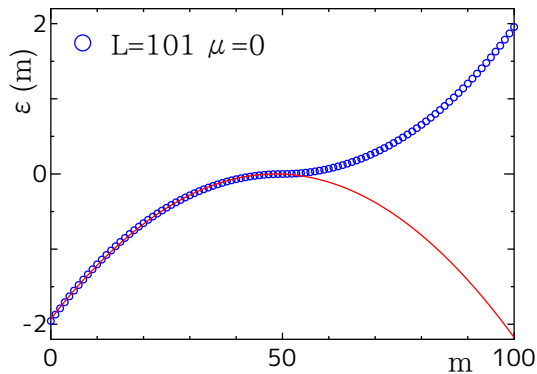}
\caption{Eigenvalue spectrum of the lattice SSD model with $L=101$ and $\mu=0$. 
The energy eigenvalues $\epsilon (m)$ are arranged in ascending order. 
The Fermi level corresponds to $\epsilon (50)$, which is exactly at zero energy.
The solid curve indicates 
$\epsilon (m) = -8.2 \times (m-49)(m-50)$
which is inferred from the spectrum of the continuous SSD analytically obtained in section 3.}
\label{fig1}
\end{center}
\end{figure}

To verify Eq. (\ref{CLcomparison}), we numerically diagonalize the lattice Hamiltonian (\ref{latticeSSD}) for $L=101$ with $\mu=0$ and obtain the single-particle eigenfunctions $\chi_m (n)$. 
In this situation, we have $51$ energy levels below the zero energy: $m=0, \cdots, 50$, where the exact zero energy state corresponds to $m=50$.
This can be seen in Fig. \ref{fig1}, where the eigenvalue spectrum is shown as a function of $m$. 
Note that,  because of the finite-$L$ effect, the energy of the $m=49$ state slightly deviates from the zero energy to the low-energy side.
Thus, the many-body ground state of the lattice SSD problem can be obtained by filling the levels $m=0, \cdots, 49$ with $N=50$ fermions.

Before proceeding to the direct comparison  of wavefunctions, we discuss the eigenvalue spectrum.
Figure \ref{fig1} shows the eigenvalue spectrum $\epsilon(m)$ for the lattice SSD problem with $L=101$ and $\mu=0$ in increasing order.
In the figure, we also plot the spectrum of the continuous SSD in the negative energy region: $\epsilon (m)= -\kappa\beta(\beta-1)$ with $\beta=50-m$. 
The overall coefficient $\kappa= 8.2 \times 10^{-4}$ is determined so that the lattice and continuous results are on top of each other. 
Note that this value of $\kappa$ is clearly consistent with  that of the continuous system $ \kappa=(\frac{\pi}{L})^2 \simeq 9.67 \times 10^{-4} $ for $L=101$. 
As depicted in the figure, we can see that the spectrum of the lattice SSD is well described by that of the continuous SSD in the entire energy range below the Fermi level.
This should be contrasted to the fact that the linearization of the single-particle dispersion of the uniform lattice system, which is essential for taking the continuum limit, is valid only in the vicinity of the Fermi level. 
In this sense,  the SSD for the lattice system efficiently smears out the lattice effect. 

Using the above correspondence between the lattice and continuous SSD, we can further explain the various characteristics of the spectrum around the Fermi energy ($\epsilon=0$) reported in Ref.~\cite{Hotta}.
For instance,   Eq. (\ref{SSDspectrum}) for a fixed $\beta(=N-m)$ leads to  
\begin{equation}
\epsilon_L = -\frac{C}{L^2},
\label{FSS}
\end{equation}
with $C := \pi^2\beta(\beta-1)$, which is in good agreement with the size dependence found in Ref.~\cite{Hotta}.
Note that this scaling relation contains no higher order correction of $O(L^{-3})$.
Also from Eq. (\ref{SSDspectrum}), we obtain the density of state (DOS) around the Fermi level as  
\begin{equation}
D(\epsilon) = \left|\frac{d \beta}{d\epsilon}\right| =\left(\frac{L}{\pi}\right)^2\frac{1}{2\beta-1}\simeq \frac{L}{2\pi} \frac{1}{\sqrt{|\epsilon|}}
\end{equation}
for $|\epsilon|  \gg 1/L^2$. 
For a sufficiently long chain, the DOS diverges with $|\epsilon|^{-1/2}$, which is also consistent with the numerical result in Ref.~\cite{Hotta}. 
For $|\epsilon| \ll 1/L^2$, however, $D \simeq (\frac{L}{\pi})^{2}$(=const), which is a cutoff bound in the vicinity of the Fermi level.

\begin{figure}
\begin{center}
\includegraphics[width=6cm]{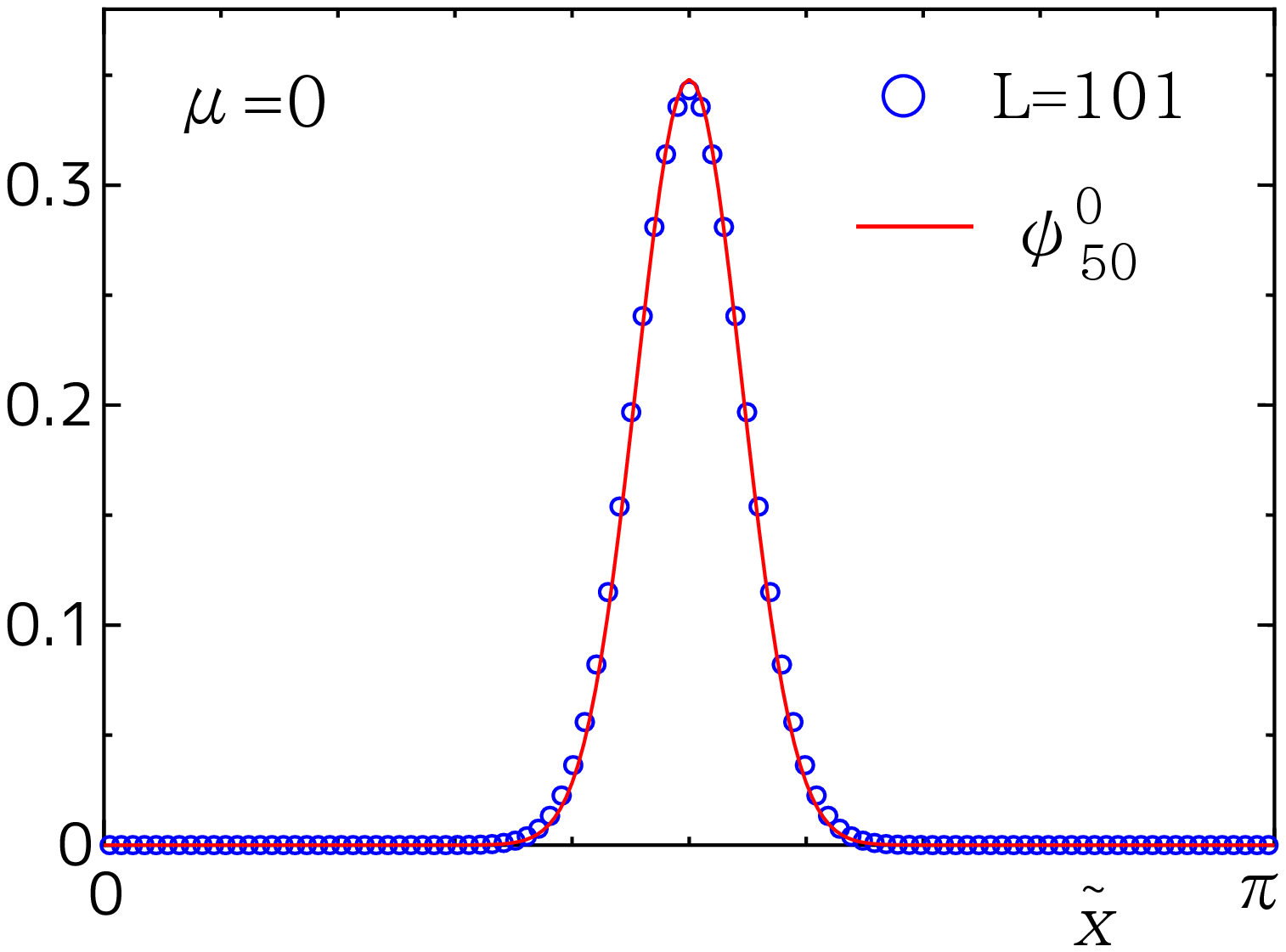} 
\includegraphics[width=6cm]{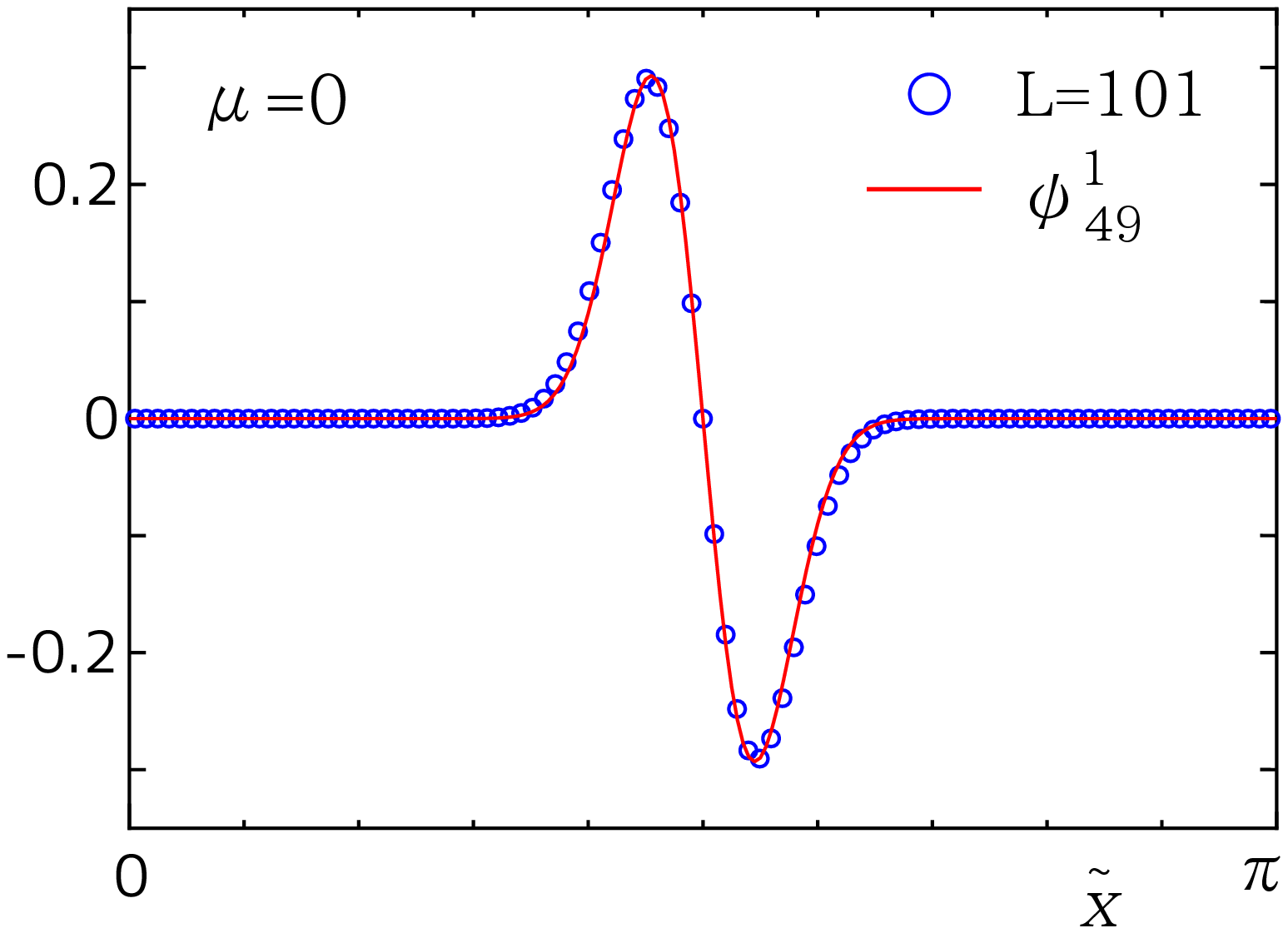} 
\includegraphics[width=6cm]{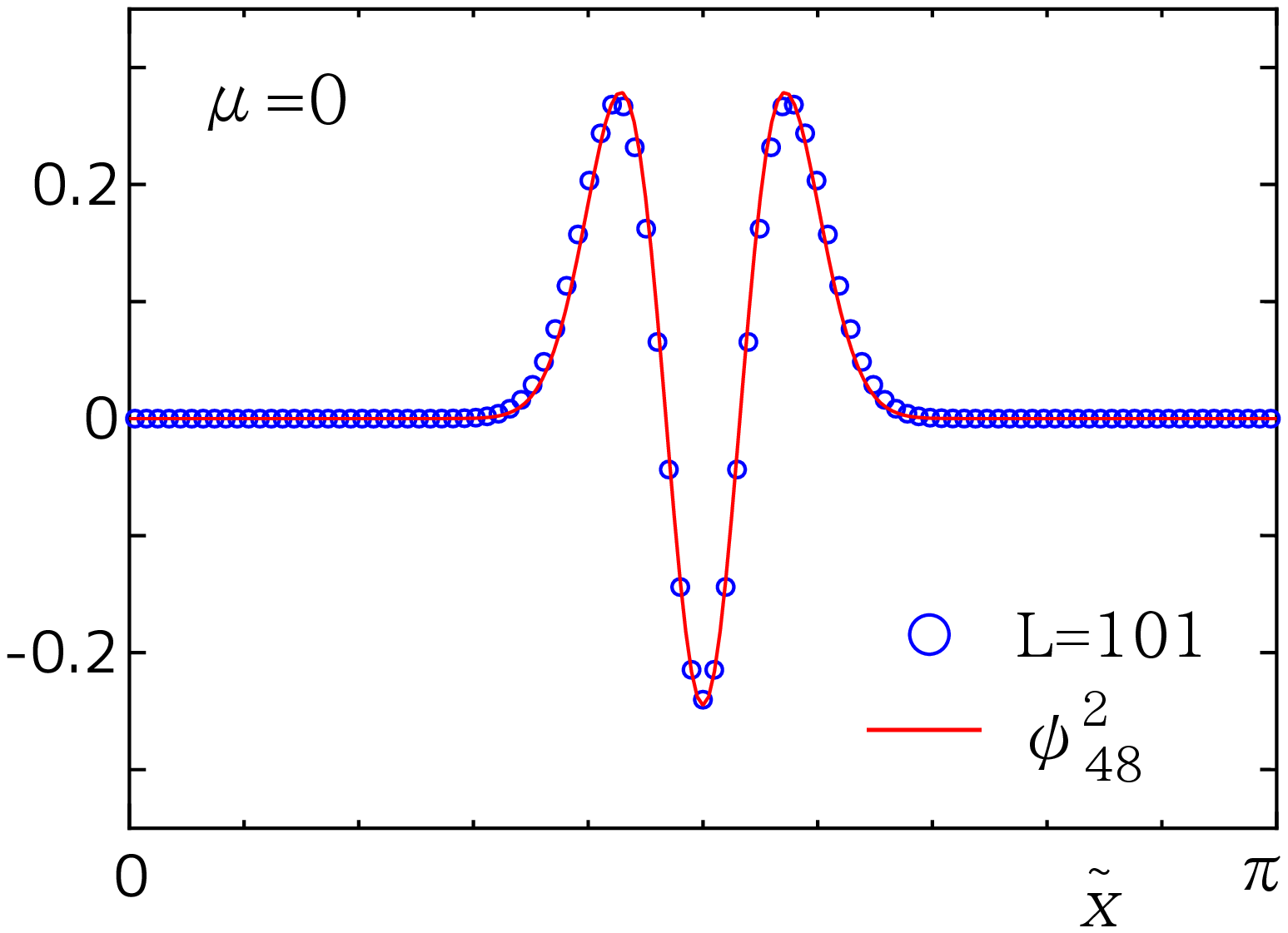} 
\includegraphics[width=6cm]{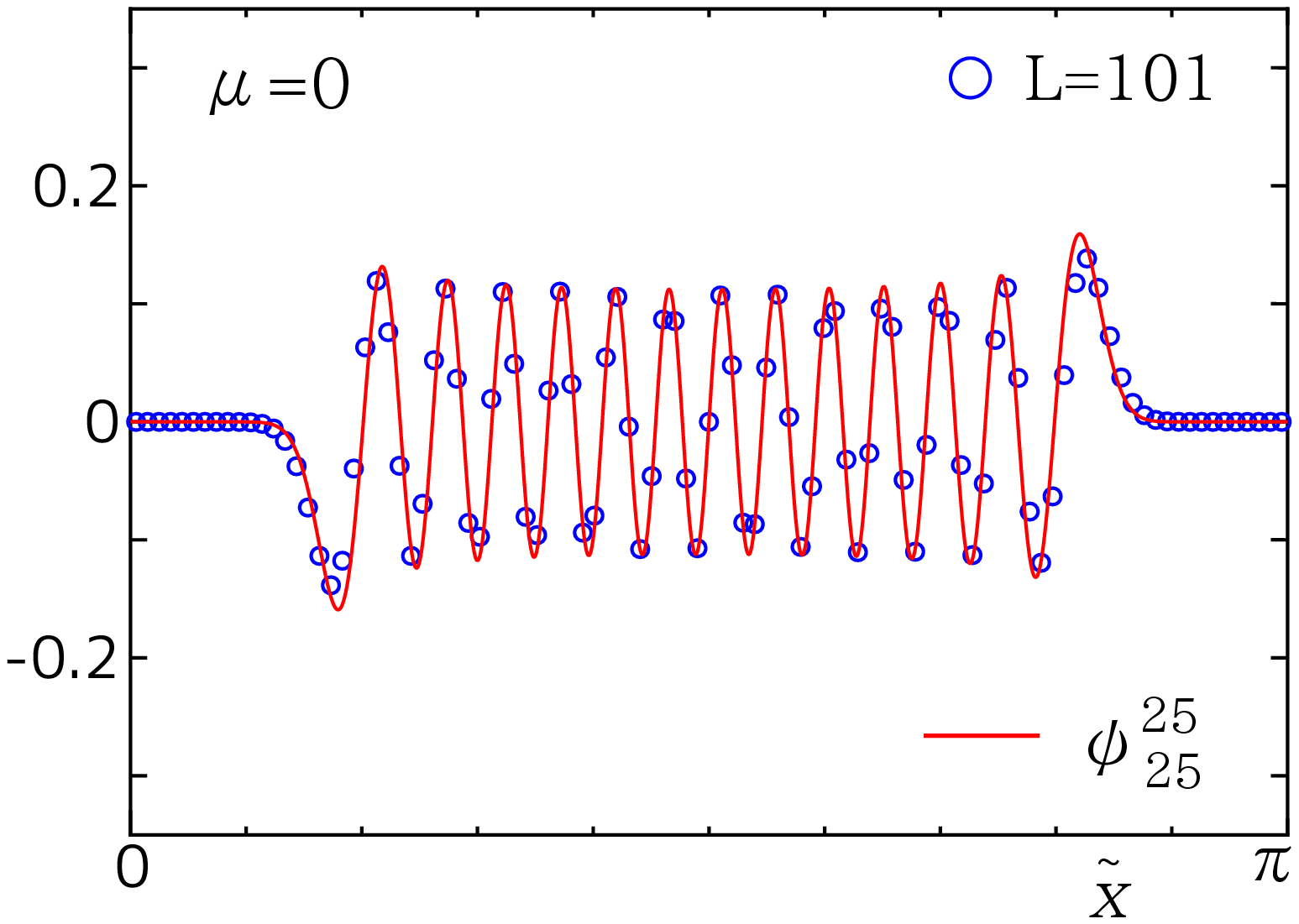}
\caption{Comparison of single-particle wavefunctions between the lattice and continuous SSD systems. The chain length and the chemical potential are set to $L=101$ and $\mu=0$, respectively. 
In the corresponding continuous SSD, the total number of fermions is fixed as $N=50$. 
The lattice wavefunctions with the geometric correction ($\psi^{\rm ssd}_m (n)$) are plotted as open circles for $m=0, 1, 2, 25$, while the continuum wavefunctions ($\psi^{m}_{N-m}(x)$) are shown as solid lines.
Note that the scaled position $\tilde{x} := \frac{\pi}{L}(n-\frac{1}{2})$ is used for $\psi^{\rm ssd}_m (n)$ for comparison.
}
\label{figwf}
\end{center}
\end{figure}

Now, we turn to the comparison of the wavefunctions between the lattice and continuous SSD problems for $L=101$ and $N=50$.
On the basis of Eq. (\ref{CLcomparison}), we define the lattice wavefunction with the geometric correction as 
\begin{equation}
\psi^{\rm ssd}_m (n):=\sin(\tilde{x})\, \chi_m (n)
\end{equation}
with $\tilde{x}= \frac{\pi}{L}(n-\frac{1}{2})$. 
The spatial profiles of $\psi^{\rm ssd}_m (n)$ for $m=0, 1, 2, 25$ are shown in Fig. \ref{figwf}.
The corresponding wavefunctions of the continuous SSD, $\psi_{N-m}^{m}(\tilde{x})$, for $m=0,1,2,25$ are also plotted as solid lines, where the overall normalization is adjusted so that $\psi^{\rm ssd}_m (n)$ and $\psi^{m}_{N-m}(x)$ are almost on top of each other. 
One can clearly see that the single-particle ground-state ($m=N-\beta=0$) is localized at the center of the chain and $\psi^{\rm ssd}_m (n)$ tends to delocalize as $m$ increases.
It is remarkable that $\psi^{m}_{N-m}(x)$ well reproduces $\psi^{\rm ssd}(n)$ even when $m=25$, which exhibits a rapid oscillation as a function of ${\tilde x}$. 
Here, we note that the geometric correction  becomes effective near the boundaries of the system.
If we omit this correction, the lattice wavefunction significantly deviates from $\psi^{m}_{N-m}(x)$, as $\tilde{x}\to 0$ or $\pi$.
We thus conclude that the correspondence (\ref{CLcomparison}) between lattice and continuous SSD is well established not only at the level of energy eigenvalues, but also at the level of individual eigenfunctions. 
This implies that the continuum limit of the lattice is taken efficiently in the SSD system, which is consistent with the fact that the leading finite-size dependence of the excitation energy is proportional to $1/L^2$, instead of $1/L$ which is universal in CFT. 

\begin{figure}[bth]
\begin{center}
\includegraphics[width=8cm]{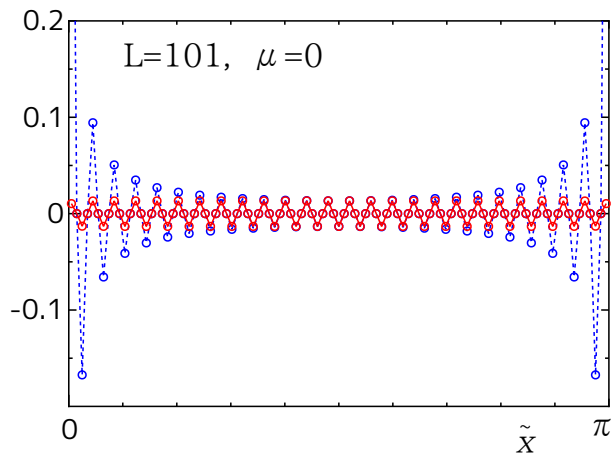}
\end{center}
\caption{The lattice SSD wavefunction at the Fermi level and geometric correction for the case of $L=101$ and $\mu=0$. 
The blue broken line with open circles indicates $\chi_{50} (n)$ as a function of the scaled position $\tilde{x}=\frac{\pi}{L}(n-\frac{1}{2})$.
The red solid line with open circles represents $\psi^{\rm ssd}_{50} (n) = \sin(\tilde{x}) \chi_{50} (n)$.}
\label{zeroewf}
\end{figure}

The geometric correction significantly affects the wavefunctions near the Fermi level, which is extended over the entire region of the chain.
In Fig. \ref{zeroewf}, we show the lattice wavefunction $\chi_{50}(n)$ for $L=101$ with $\mu=0$, where $m=50$ corresponds to the zero energy state just on the Fermi level.
In the figure, we also plot $\psi^{\rm ssd}_{50} (n) \propto \psi^{50}_0 (x)$ for comparison.
Then, it is observed that the amplitude of the lattice wavefunction  $\chi_{50} (n)$ increases toward the end sites $n=1$ and $L$, while $\psi^{\rm ssd}_{50} (n) \sim \psi_{0}^{50}(\tilde{x})$ retains a plane-wave behavior. 
As was mentioned before,  $\psi_0^N(x)$ is not normalizable with respect to the inner product Eq. (\ref{innerprod}) since $\psi_0^N(x)/\sin (N \tilde{x}) \sim \cot(N\tilde{x})$  diverges at $\tilde{x}=0$ and $\pi$.
This divergence comes from the fact that $\psi_0^N(x)$ has a finite amplitude at the edges due to the Neumann boundary conditions. 
The result of Fig. \ref{zeroewf} 
implies that the divergence of the inner product (\ref{innerprod}) for $\psi_0^N$, which originates solely from the geometric factor, can be regularized by a lattice cutoff.

\section{summary and discussions}

In this paper, we have studied the sine-square deformation (SSD) of the free-fermion problem in one-dimensional continuous space.
We formulated the single-particle problem of the SSD system as the inverse of the $1/\sin^2$ potential problem, 
where supersymmetric (SUSY) quantum mechanics is very effective.
Exploiting both SUSY and the shape invariance, we explicitly obtained the complete set of the single-particle eigenfunctions of the SSD problem, and showed that its linear space is identical to that of the uniform periodic (antiperiodic) ring. 
Accordingly, we have shown that the many-body ground state of the system with SSD is the same as that of the uniform system with periodic (antiperiodic) boundary conditions. 
It would be interesting to generalize our theory to other class of superpotentials with the shape invariance. 
In addition, we have found that various properties of the lattice SSD problem are well explained by the analytic results for the SSD in the continuous space.  
In particular, the eigenfunctions of the lattice SSD problem can be reproduced from those of the continuous SSD by taking into account the geometric correction.

In the previous studies of the lattice systems that are reducible to free fermions~\cite{Katsura2011, Katsura2011B, MaruyamaKH2011}, the ground-state correspondence was proved using the plane-wave basis. 
Also, the correspondence was explained in the context of CFT~\cite{Katsura2011B}, where the SSD Hamiltonian is expressed as $L_0-(L_{+1}+L_{-1})/2$ and both the Virasoro generators $L_\pm$ annihilate the vacuum state $|0\rangle$.
The present study clearly provides a complementary approach to the SSD problem, and various aspects of the SSD eigenfunctions can be clarified with the help of SUSY quantum mechanics.
For example, we have demonstrated in the SSD system that the  $1/L^2$  correction (\ref{FSS}) of the excitation energy emerges with no higher order terms, instead of the universal $1/L$-size dependence in CFT. 
In this paper, nevertheless, SUSY quantum mechanics was just 
used to systematically construct the single-particle eigenfunctions of the SSD problem, and thus the role of the SUSY may not be clear at the level of field theory.
Moreover, a very recent paper~\cite{Ishibashi} pointed out that an infinite circumference limit of CFT can be obtained by adopting $L_0-(L_{+1}+L_{-1})/2$ as the Hamiltonian instead of $L_0$.  
Such singular behavior may be related to the normalizability of the zero energy state satisfying the Neumann boundary conditions that appear in the present continuous SSD.
Therefore, it would be interesting to explore the relation between SUSY and CFT behind the SSD problems. 

\ack{}
The authors would like to thank 
T. Hikihara, T. Nishino, K. Ohta, and  T. Tada 
for useful discussions and comments.
This work was supported in part by Grants-in-Aid Nos. 26400387, 25400407, and 15K17719 from the Ministry of Education, Culture, Sports, Science and Technology of Japan. 

\appendix

\section{Formal solutions for positive energy}
\label{app: positive_eng}

\subsection{$0< \xi < \frac{1}{4}\left(\frac{\pi}{L}\right)^2 $}

For Eq. (\ref{SSDpotential}),  $\xi$ becomes positive in $0<\beta<1$, which could be a candidate of a scattering state with positive energy.
Taking account of a non-integer valued $\beta$, we introduce another parameterization $\beta=a+1/2$, which yields   
\begin{equation}
\xi = -\left(\frac{\pi}{L}\right)^2 \left( a^2-\frac{1}{4} \right). 
\label{epsilon_a}
\end{equation}
We then obtain  the linearly-independent solutions  of Eq. (\ref{SSDpotential}) for a fixed $\mu = (\frac{\pi}{L})^2 q^2$ as
\begin{eqnarray}
u(x)  &=& (\sin \tilde{x})^{1/2 + a} \cos \tilde{x}\,
F \left(
\frac{3}{4} + \frac{a}{2} - \frac{q}{2},\, 
\frac{3}{4} + \frac{a}{2} + \frac{q}{2};\, 
\frac{3}{2};\, \cos^2 \tilde{x} \right)\label{hypergeo1}
  \\
v(x) &=&  (\sin \tilde{x})^{1/2 + a}\, 
F \left(
\frac{1}{4} + \frac{a}{2} - \frac{q}{2},\, 
\frac{1}{4} + \frac{a}{2} + \frac{q}{2};\, 
\frac{1}{2};\, \cos^2 \tilde{x} \right),
\label{hypergeo2}
\end{eqnarray}
where $F$ is Gauss's hypergeometric function. 
Note that, if $a$ is a positive half integer, Eqs. (\ref{hypergeo1}) and (\ref{hypergeo2}) reduce to the Gegenbauer polynomials.
Otherwise, these are not a polynomials in $\cos^2 {\tilde x}$. 

Since Eqs. (\ref{hypergeo1}) and (\ref{hypergeo2}) with $|a|<1/2$ are not divergent in $0\le x \le L$, one might expect that the scattering state could be described by a linear combination of them.
However, the derivative of $u(x)$ and $v(x)$ with respect to $x$ 
diverges at $\tilde{x}=0$ and $1$ when $|a|<1/2$, as far as $q$ is fixed at an integer~\cite{Scarf}.
Thus, there is no physical solution when $0 < \xi < \frac{1}{4}(\frac{\pi}{L})^2$. 
Of course, in the limit of $a\to \pm 1/2$, $u(x)$ and $v(x)$ approach to the trigonometric functions that are solutions for $\beta=0$ and $\beta=1$.

\subsection{$\xi > \frac{1}{4}\left(\frac{\pi}{L}\right)^2$}

In the context of the $1/\sin^2$ potential problem, the coupling corresponding to $\xi > \frac{1}{4}\left(\frac{\pi}{L}\right)^2$ is so strongly attractive that the single-particle ground state of $H_\beta$ is unstable, i.e., $\mu$ is not bounded from below. 
Formal solutions of Eq. (\ref{SSDpotential}) for a certain fixed chemical potential $\mu$ are also given by Gauss's hypergeometric function
\begin{eqnarray}
u(x)&=&(\sin \tilde{x})^{1/2 + ia}\cos \tilde{x}\,
F \left(
\frac{3}{4} +i \frac{a}{2} - \frac{\mu}{2},\, 
\frac{3}{4} +i \frac{a}{2} + \frac{\mu}{2};\, 
\frac{3}{2};\, \cos^2 \tilde{x} \right) 
\label{hypergeo3}\\
v(x)&=&(\sin \tilde{x})^{1/2 + ia}\,
F \left(
\frac{1}{4} + i\frac{a}{2} - \frac{\mu}{2},\, 
\frac{1}{4} + i\frac{a}{2} + \frac{\mu}{2};\, 
\frac{1}{2};\, \cos^2 \tilde{x} \right),
\label{hypergeo4}
\end{eqnarray}
where we have adopted the parameterization $\xi = \left(\frac{\pi}{L}\right)^2 (a^2+\frac{1}{4})$. 
However, Eqs. (\ref{hypergeo3}) and (\ref{hypergeo4}) exhibit singular behavior in the vicinities of $\tilde{x}=0$ and $1$, reflecting the unstable ground state.

\section{Derivation of $I_{N,\beta}$ in Eq. (\ref{innerprod})}
\label{app:coeffI}
In this Appendix, we present a detailed derivation of the coefficient $I_{N,\beta}$ appearing in Eq. (\ref{innerprod}). 
Consider the squared norm of $\psi^{\ell}_{\beta} (x)$, 
where $\beta$ is assumed to be an integer and $\beta \ge 1$. 
Changing the variable from $x$ to $z:= \cos \frac{\pi}{L}x$, we have
\begin{equation}
\langle\langle \psi^{\ell}_{\beta},\, \psi^{\ell}_{\beta} \rangle\rangle
=\frac{({\cal N}^{\ell}_{\beta})^2}{\pi} 
\int^1_{-1} dz\, [C^{\beta}_{\ell} (z)]^2\, (1-z^2)^{\beta-\frac{3}{2}}.
\end{equation}
To evaluate the above integral, we expand the Gegenbauer polynomials $C^{\beta}_\ell$ in terms of $C^{\beta-1}_{n}$ ($n=0,1,...,\ell$). The results depend on whether $\ell$ is even or odd:
\begin{equation}
C^{\beta}_{2m} (z) = \sum^m_{k=0} \frac{\beta+2k-1}{\beta-1} C^{\beta-1}_{2k} (z),
\quad
C^{\beta}_{2m+1} (z) = \sum^m_{k=0} \frac{\beta+2k}{\beta-1} C^{\beta-1}_{2k+1} (z),
\end{equation}
which can be proved using the formula obtained by Gegenbauer himself \cite{Keiner}. 
Then, using the orthogonality of $C^{\beta-1}_\ell (z)$:
\begin{equation}
\int^1_{-1} dz\, C^{\beta-1}_{\ell} (z) C^{\beta-1}_{\ell'} (z) (1-z^2)^{\beta-3/2}
=\frac{\pi\, 2^{3-2\beta}\, (2\beta+\ell-3)!}{(\beta+\ell-1)\, \ell!\,  [(\beta-2)!]^2}\,
\delta_{\ell,\ell'},
\end{equation}
we arrive at
\begin{equation}
\langle\langle \psi^{\ell}_{\beta} ,\, \psi^{\ell}_{\beta} \rangle\rangle
=2^{2-2\beta}\, ({\cal N}^{\ell}_{\beta})^2\,
\frac{(2\beta+\ell-1)!}{(2\beta-1)\, \ell!\, [(\beta-1)!]^2}.
\end{equation}
The coefficient $I_{N,\beta}$ can be read off from the above formula by replacing $\ell$ with $N-\beta$.

\section*{References}
\providecommand{\newblock}{}


\begin{thebibliography}{99}
\bibitem{chiral} The meaning of ``chiral'' in this context is different from the original meaning in particle physics and field theory.
\end{thebibliography}
\end{document}